\documentstyle[preprint,eqsecnum,aps,epsfig]{revtex}

\newcommand{\be}{\begin{equation}}   
\newcommand{\ee}{\end{equation}}
\newcommand{\beq}{\begin{eqnarray}}
\newcommand{\eeq}{\end{eqnarray}}
\newcommand{\beqn}{\begin{eqnarray*}}
\newcommand{\eeqn}{\end{eqnarray*}}
\newcommand{\f}[2]{\frac{#1}{#2}}

\def\ms{M_\star}
\def\mp{M_p}
\def\op{ \ $ }
\def\cl{$ \ }
\def\nn{\nonumber}

\def\f{\frac}

\def\Y{Y_{l m}}

\def\EM{e^{-i\omega t}}

\def\N{N_{l m}}
\def\T{T_{l m}}
\def\V{V_{l m}}
\def\L{L_{l m}}
\def\RN{N_{l m}(\omega,r)}
\def\RT{T_{l m}(\omega,r)}
\def\RV{V_{l m}(\omega,r)}
\def\RL{L_{l m}(\omega,r)}

\begin{document}
\draft

\title{Gravitational waves emitted by solar-type stars
excited by  orbiting planets}

\title{Excitation  of g-modes of solar type stars by an orbiting
companion}

\author
{E. Berti and V.  Ferrari}

\address
{ Dipartimento di Fisica ``G.Marconi",
 Universit\` a di Roma ``La Sapienza"\\
and \\
Sezione INFN  ROMA1, p.le A.  Moro
2, I-00185 Roma, Italy}

\date{\today}

\maketitle

\begin{abstract}
The possibility of exciting the g-modes of a solar-type star as 
a consequence of the gravitational interaction with a close companion 
(a planet or a brown dwarf) is studied by a perturbative approach.
The amplitude of the emitted gravitational wave is computed and compared
with the quadrupole emission of the system, showing that in some cases
it can be considerably larger. 
The effects of radiation reaction are considered 
to evaluate the timescale of the emission process, and a Roche lobe
analysis is used to establish the region where the companion can orbit without
being disrupted by tidal interactions with the star.

\end{abstract}


\narrowtext
\section{Introduction}

In recent years  a significant number of systems composed by a 
solar-type star and one or
more orbiting companions has been discovered in our neighbourhood.
According to the {\it Extrasolar Planets Catalog}\cite{Uff2},
at the time we write this paper  58 extrasolar planetary systems
have been confirmed, and this number is destinated to grow in the near
future (see \cite{review} for a recent review).
46 of the observed orbiting objects are planets with masses ranging
between [0.16-11]$M_J$, where $M_J$ is Jupiter's mass; 12 are bigger, up
to  $60~M_J$, and include  super-planets and brown dwarfs.
In the following, we shall indicate all the orbiting objects
as `planets'. 

The newly discovered systems exhibit some unexpected features; 
22 planets  move on elliptic orbits with eccentricities $e > 0.3$,
larger than the largest  eccentricity encountered in  the solar system
(Mercury, $e = 0.2$), and  many of them are incredibly
close to the central star, much closer than expected. 
For instance the planet orbiting  51 Pegasi, discovered in
1995\cite{51Pega},\cite{51Pegb}, has a period $P=4.23$ days 
and orbits its star at a
distance  $a=0.05$ AU.  
More than two thirds of the planets discovered up to now are orbiting their
host star much closer than Mercury orbits the Sun
($P=88$ days,   $a=0.39$ AU), and even closer than 
51 Pegasi's planet.
These orbital properties  challenge the existing theories on planet
formation and evolution, and suggest an interesting possibility:
there may exist planets on  orbital radii sufficiently small to excite 
some g-modes of the star by tidal interaction.
This issue has been discussed by Terquem et al. \cite{terquem}
in connection with the short orbital period of 51 Pegasi's planet.  
They calculated the dynamical tides raised by the planet on the star, and
the dissipation timescales due to turbulent viscosity in the
convective zone and to radiative damping in the radiative core.
In this paper we shall approach the problem of the excitation of the
g-modes from a different point of view: by using a perturbative 
approach,
justified by the fact that the mass of the central star is much bigger
than that of the planet, we shall compute the 
gravitational signal emitted by a  system in which the planet  moves on
a circular orbit of radius ${R}_0$, close to that which would correspond 
to the resonant excitation of a g-mode,  $R_{res}$. 
We shall extend the work presented in a previous paper
(\cite{bdf}, to be referred to hereafter as Paper I),
were we have studied the quadrupole emission due to the orbital motion
of the  observed planetary systems,
and started to  explore the  possibility for a planet
to excite  some stellar modes.
First of all, in section II we shall verify by a Roche-lobe analysis whether
the planet can move on the orbit  $R=R_0$ without being disrupted by tidal
interactions.
In section  III,  using the quadrupole formalism we shall compute
the characteristic amplitude of the gravitational signal, $h_{Q}$,
emitted by the system because of the orbital motion  of
the planet  on  ${R}_0$. As usual, the two bodies will
be treated as pointlike masses, and the effects of the planet on the
internal structure of the star will be neglected.
These effects will be considered in section IV,
where  we shall assume that the 
planet induces a small perturbation on the gravitational
field of the star and on its thermodynamical structure.  
We shall  integrate 
the perturbed Einstein's equations coupled to  the hydrodynamical equations,
having the stress energy tensor of the planet as a source,  and compute
the characteristic amplitude of the emitted
gravitational signal, $h_{R}$, which will be compared with $h_{Q}$.
In this way we will be able to evaluate the enhancement in the radiation
emitted by the system as a consequence of the g-modes excitation.
In Section V we shall discuss how long can a planet  move on 
an orbit close to resonance before radiation reaction effects
lead the planet off resonance. 
Conclusions will be drawn in section VI.

\section{Which stellar modes can be excited by an orbiting planet}
In this section we shall verify, by a Roche-lobe analysis,
if a planet can orbit  on a radius such that  
the keplerian orbital frequency, $\omega_k=\sqrt{G(\ms+\mp)/R_0^3},$ 
is half the frequency of a given mode of  the star
$\omega_i$. 
As discussed in \cite{terquem},\cite{alexander} and \cite{shaferkostas},
in this case the mode would be excited by the dynamical tides raised by the 
planet.
We shall consider as a model  a polytropic 
star, \op p = K\epsilon^{1+1/n}$
with $n=3,\cl $\epsilon_0/p_0=5.53\cdot 10^5$
 and adiabatic exponent $\gamma=5/3$. Choosing 
the central density \op\epsilon_0=76$ g/cm$^3$,
this model gives a star with the  same mass and radius as the Sun.
In Paper I we showed that if one considers a polytropic star with $n=2$
and mass equal to that of the Sun,  the p-modes and
the fundamental mode cannot be excited by a planet, because it would be
tidally disrupted. The same holds for the more appropriate, though still
simplified, $n=3$ model considered in this
paper, therefore we shall deal only with the excitation of the g-modes.

We shall consider three companions for the central star: two planets with
the mass  of the Earth and of Jupiter,  and a
brown dwarf  of 40 jovian masses,
$M_E,$ $M_J$ and $M_{BD},$ respectively.
Smaller planets produce gravitational signals that are too small to be 
interesting.
Let us assume that a planet orbits the star at a  constant orbital radius
$R^i_{res}$, in a position  to  excite the mode $g_i$.
Following the procedure described in Paper I it is possible to determine,
by a Roche lobe analysis, what is the maximum radius  the planet 
can have in order not to overflow its Roche lobe.
Since we have fixed the mass of the planet,  this constraint also
determines  the minimum value of the planet's mean density, $\rho_{min}$,
compatible with the excitation of that mode.
The results of this analysis are shown in table 1,
where  we tabulate the value of $\rho_{min}$, 
expressed in units of the
mean density of the Sun, for the three considered  companions.
It should be stressed that in the Roche lobe analysis
the ratio $\rho_{min}/\rho_\star$, for each given mode,
depends only on the ratio between the
mass of the planet and that of the star.
From table \ref{lobi} we see that, since 
$\rho_{Earth}/\rho_\odot=3.9\cl and $\rho_{J}/\rho_\odot=0.9,\cl
a planet like the Earth can orbit sufficiently close to the star to
excite g-modes of order higher or equal to $n=4$, whereas a planet like Jupiter 
can excite only the mode $g_{10}$ or higher.

According to  the brown dwarf model (``model G'') by Burrows
and Liebert \cite{burrows} an evolved, 40 $M_J$
brown dwarf has a radius $R_{BD}=5.9\cdot 10^4$ km, and a corresponding
mean density $\rho_{BD}=88$ g$\cdot$cm$^{-3}$; consequently
$\rho_{BD}/\rho_\odot=64$, and this value is high enough to
allow a brown dwarf companion to excite all the g-modes of the central
star.  However,  we also need to take into account
the destabilizing mechanism  of  mass
accretion {\it from} the central star. This imposes a further constraint,
and this is the reason why the slots corresponding to the 
excitation of the g-modes lower  that g$_4$ in the last row of table
\ref{lobi} are empty.

\section{The quadrupole emission of a binary system}
In this section we shall use the quadrupole
formalism to  compute the gravitational signal emitted by
a binary system.
We shall assume that a planet orbits its host star in circular motion,
and that the radius of the orbit corresponds to the excitation of one of
the g-modes allowed by the Roche lobe analysis discussed in section II.
As usual, both the star and the planet will be treated as pointlike
masses, with no reference to their internal structure.
From the reduced quadrupole moment 
\be
Q_{kl}=
\mu \left(X^{k}X^{l}-\frac{1}{3}\delta^{k}_{l}\left|{\mathbf{X}}
\right|^{2}\right),
\ee
where   $\mu = M_p M_\star/(M_p+M_\star)$  and
 \op {\mathbf{X}}={\mathbf{x}_\star}- {\mathbf{x}_p}\cl is the relative
position vector,
the non vanishing TT-components of the emitted wave  can easily be
computed
\be
\label{wave}
h_{ij}^{TT}(t,\ {\mathbf{x}})  =  \frac{2}{r} \left(
P_{ik}P_{jl}-\frac{ 1}{2}P_{ij}P_{kl} \right)
\ddot{Q}_{kl}(\tau)|_{\tau=t-\frac{r}{c}}
\ee
where \op P_{ik}=\delta^{i}_{k}-n_{i}n_{k}\cl is the projector
onto the 2-sphere \op r=const,\cl and ${\mathbf{n}}$ is the radial
unit vector.
It should be stressed that  eq. (\ref{wave}) allows to compute 
the radiative contribution due solely to the {\it orbital motion} of the system.
For a circular orbit 
the wave is emitted at twice the keplerian frequency
\cite{PetersMathews}, and
from the two independent wave components \op h_{\pm}
(2\omega_k,r,\vartheta,\varphi)\cl
we compute the characteristic amplitude \cite{Kip}
\be
\label{hc}
h_Q(2 \omega_k,r)=\sqrt{\frac{2}{3}}
\sqrt{\f{1}{4\pi}\int{d\Omega
\left[\left|{h}_+\right|^2
+\left|{h}_-\right|^2\right]}}=
\f{M_p}{r}4\sqrt{\f{2}{15}}
\left(\omega_k R_0\right)^2,
\ee
where the factor $\sqrt{2/3}$ takes into account the average over
orientation, and \op\Omega\cl is the solid angle.
This amplitude will be compared with that  determined by the 
perturbative approach.
In table \ref{amplitudes} we give the values of
$h_Q\cl for the three companions
orbiting the sun-like star which we use as a model. 
The orbital radius is chosen to
correspond to the excitation of a permitted g-mode
(cfr. table \ref{lobi}).
The planetary systems are assumed to be located at a distance of
$10$ pc from Earth.
In the last row  we give the frequency (in $\mu$Hz) of the g-modes,
which is also the frequency at which waves are emitted.

\section{The gravitational signal emitted by a perturbed star}
In this section we shall briefly introduce the equations of general relativity
which describe the perturbations of a  star excited by a planet
moving on a circular orbit.
Unlike the quadrupole formalism discussed in section III, the perturbative
approach takes into account the internal structure of the star and the
modification induced on it by the interaction with the planet, and it allows
to evidentiate the contribution to the emitted wave
due to  the excitation of the stellar modes.
When the orbit is circular, only the polar, nonaxisymmetric perturbations
are  excited, and the metric  appropriate for their description can be
written in the following form
\cite{ferrarigualtieribor}
\beq
\label{pertmetric}
ds^2 &=&
e^{2\nu(r)} dt^2 - e^{2\mu_2(r)} dr^2 -
r^2 d\vartheta^2-r^2\sin^2\vartheta d\varphi^2\\
\nn
&+&2\sum_{l m}~~\int^{+\infty}_{-\infty} d\omega~\EM
\Biggl\{
 \left[
e^{2\nu}\N dt^2- e^{2\mu_2}\L dr^2\right]\Y -  r^2 \left[\T+ 
\V\frac{\partial^2}{\partial\vartheta^2}
\right]\Y d\vartheta^2  \Biggr.\\
\nn
&-&\Biggl.
 r^2\sin^2\vartheta
\left[\T+ \V \left( \frac{1}{\sin^2{\vartheta}}
     \frac{\partial^{2}}{\partial \varphi^{2}}
     +\cot{\vartheta}\frac{\partial}{\partial \vartheta}
     \right) \right] \Y  d\varphi^2
\Biggr.\\
\nn
&-&\Biggl. 2r^2 \V\left[
\frac{\partial^2}{\partial\varphi\partial\vartheta}
-\frac{\partial}{\partial\varphi}\cot\vartheta\right]\Y
 d\vartheta d\varphi
\Biggr\}~,
\eeq
where the functions $[\RN,\RT,\RV,\RL]$ describe the radial part of the
perturbed metric, and $Y_{l m}(\vartheta,\varphi)$ are the scalar 
spherical harmonics.
The metric functions $\nu(r)$ and $\mu_2(r)$ describe the
unperturbed spacetime and are  determined by
numerically integrating the TOV equations  for hydrostatic
equilibrium for the chosen equation of state. 
The functions $[N,T,V,L]$
have to be found by solving the perturbed Einstein equations outside the
star and inside, where they couple to the hydrodynamical equations. 
Inside the star, after separating the variables and assuming a 
nonbarotropic equation of
state, the perturbed equations can be reduced to the 
following set, from which the
hydrodynamical variables have been eliminated \cite{ferrarigermano}
\beq
\label{poleq}
&&X_{,r,r}+\left(\frac{2}{r}+\nu_{,r}-\mu_{2,r}\right)X_{,r}+
\frac{n}{r^{2}}
e^{2\mu_{2}}(N+L)+\omega^{2}e^{2(\mu_{2}-\nu)}X=0,
\\\nn
&&(r^{2}G){,r}=n\nu_{,r}(N-L)+\frac{n}{r}(e^{2\mu_{2}}-1)(N+L)+r(\nu_{,r}
-\mu_{2,r})X_{,r}+\omega^{2}e^{2(\mu_{2}-\nu)}rX\; ,
\\\nn
&&-\nu_{,r}N_{,r}=-G+\nu_{,r}[X_{,r}+\nu_{,r}(N-L)]+
+\frac{1}{r^{2}}(e^{2\mu_{2}}-1)(N-rX_{,r}-r^{2}G)\\
\nn
&&-e^{2\mu_{2}}(\epsilon+p)N
+\frac{1}{2}\omega^{2}e^{2(\mu_{2}-\nu)}
\left\{ N+L+\frac{r^{2}}{n}G+\frac{1}{n}[rX_{,r}+(2n+1)X]\right\},
\\\nn
&&L_{,r}(1-{D})+L\left[ \left(
\frac{2}{r}-\nu_{,r}\right)-\left(\frac{1}{r}
+\nu_{,r}\right)D\right]+X_{,r}+X\left(\frac{1}{r}-\nu_{,r}\right)+
{D}N_{,r}\\
&&+N \left( {D}\nu_{,r}
-\frac{{D}}{r}-{F}\right)+
\left(\frac{1}{r}+{E}\nu_{,r}\right)\left[ N-L+\frac{r^{2}}{n}G+
\frac{1}{n}\left(rX_{,r}+X\right)\right]=0
\; ,\nonumber
\eeq
where
\beq
\label{Ee}
{D}&=&
1-\frac{\omega^{2}e^{-2\nu}}
{2\left[\omega^{2}e^{-2\nu}+
{e^{-2\mu_{2}}\nu_{,r}(\epsilon_{,r}- Q p_{,r})}/{(\epsilon+p)}\right]},\\
{E}&=&{D}(Q-1)-Q,\nonumber\\
{F}&=& \frac{\epsilon_{,r}-Qp_{,r}}{\omega^{2}e^{-2\nu}+
{e^{-2\mu_{2}}\nu_{,r}(\epsilon_{,r}-Q p_{,r})}/{(\epsilon+p)}}~.
\nonumber
\eeq
In eqs. (\ref{poleq}) the harmonic indices \op (l,m)\cl have been suppressed,
we have replaced the function $V$ 
by $X=nV$, with \op n=(l -1)(l+2)/2$, and the function $T$ by
\beqn
G=\nu_{,r}\left[\frac{n+1}{n}X-T\right]_{,r}&+&
\frac{1}{r^{2}}(e^{2\mu_{2}}-1)[n(N+T)+N]+\frac{\nu_{,r}}{r}(N+L)-
\\
&-&e^{2\mu_{2}}(\epsilon+p)N+
\frac{1}{2}\omega^{2}e^{2(\mu_{2}-\nu)}\left[L-T+\frac{2n+1}{n}X\right]\: .
\eeqn
We have also defined  $Q=(\epsilon +p)/\gamma p,$  where
$\gamma$ is the adiabatic exponent
\[
\gamma=\frac{(\epsilon+p)}{p} \left(\frac{\partial p}{\partial\epsilon }
\right)_{entropy=const},
\]
and \op\epsilon$ and $p$ are the energy density and the pressure of the
fluid composing the star.
Since in this paper we apply the relativistic
theory to nonbarotropic stars, we choose, as in ref.
\cite{ferrarigermano}, $\gamma= const=5/3$.
The system of eqs. (\ref{poleq}) can be solved numerically by integrating
the two independent solutions which satisfy the regularity condition at the
center, and superimposing them  in such a way that the perturbation of
the pressure vanishes at the boundary, as discussed in \cite{nonradial1}.

Outside the star, \op\epsilon\cl and \op p\cl vanish,
and a source term given by the stress-energy tensor of the moving planet
must be added on the right-hand-side of the Einstein equations. 
Eqs. (\ref{poleq}) reduce to the Zerilli equation \cite{zerilli}
for the Schwarzschild perturbations
\be
\label{zereq}
\Biggl\{\frac{d^{2}}{dr_{*}^{2}}+
\omega^{2}-\frac{2(r-2\ms) 
[n^{2}(n+1)r^{3}+3\ms^{2}r^{2}+
9\ms^{2}nr+9\ms^{3}]}{r^{4}(nr+3\ms)^{2}}
\Biggr\} Z_{l m}(\omega,r)=S_{l m}(\omega,r),
\ee
where \op r_*=\int_0^re^{-\nu+\mu_2}dr,\cl and $\ms$ is the mass of the star.
The Zerilli function matches continuously with the solution of eqs.
(\ref{poleq}) in the interior, i.e., at the surface of the star
\be
\label{zerfun}
Z_{l m}(\omega, R)=\frac{R}{nR+3\ms}\Bigl(
\frac{3\ms}{n}X_{l m}(\omega,R)-RL_{l m}(\omega,R)
\Bigr),
\ee
and similarly  for its first derivative (see ref.\cite{nonradial1} for
details).  
The source term, $S_{l m}(\omega,r),$ is derived from the 
stress-energy tensor of the planet, considered as a pointlike particle
moving on a circular orbit around the star
in the equatorial plane
\be
T^{\mu\nu}=\frac{\mp}{\sqrt{-g}}\frac{dT}{d\tau}\frac{dz^{\mu}}{dt}\frac{dz^
 {\nu}}{dt}\delta(r-R_0)\delta(\theta-\pi/2)
\delta(\varphi-\omega_k t).
\ee
From the geodesic equations 
\op
\f{dT}{d\tau}=
E\left(1-\frac{2\ms}{R_0}\right)^{-1},
\cl
where the planet's energy  per unit rest mass is
\be
\label{energyandmom}
E=\left(1-\frac{2\ms}{R_0}\right)\left(1-\frac{3\ms}{R_0}\right)^{-1/2}.
\ee
In terms of these quantities the source term can be written as
\beq
\label{source}
S_{l m}(\omega,r)=
&&M_p\left\{
\f{8\pi (r-2\ms)(12\ms^2+r^2n^2+3nr\ms-6r\ms)}{\omega\sqrt{n+1} r(nr+3\ms)^2}
B^{(0)}_{l m}\right.\\
\nn
&-&\left.\f{8\pi (r-2\ms)^2}{\omega\sqrt{n+1} (nr+3\ms)} B^{(0)}_{l m~,r}
+\f{8\pi \sqrt{2}(r-2\ms)}{\sqrt{n(n+1)}} F_{l m}\right\}
\eeq
where
\beq
\label{source1}
{B}_{l m}^{(0)}(\omega,r)&=&
\sqrt{\f{2}{l(l+1)}}\f{m \omega_k}{\sqrt{1-3\ms/R_0}}
\f{r-2\ms}{r^2}
P_{l m}(\pi/2)
\delta(r-R_0)
\delta(\omega-m\omega_k).
\\\nn
{F}_{l m}(\omega,r)&=&
\f{ \left[l(l+1)-2m^2\right]}
{\sqrt{2(l-1)l(l+1)(l+2)}}
\f{\ms/R_0^3}{\sqrt{1-3\ms/R_0}}  P_{l m}(\pi/2)
\delta(r-R_0) \delta(\omega-m\omega_k).
\eeq
In order to find the amplitude of the emitted wave we 
follow the same approach used by Kojima \cite{kojima},
who first calculated the gravitational signal emitted by a particle in
circular motion around a star. In particular,
he considered the excitation of the fundamental mode of a
neutron star by an orbiting particle, showing that
a sharp resonance occurs if the frequency
of the f-mode  is twice the orbital  frequency,
and that the characteristic wave amplitude emitted at that frequency
can be up to 100 times larger than that evaluated by the quadrupole formula.
As discussed in section II, if the central star is of solar-type
only the  g-modes can be excited by  a  planet, and
therefere we shall focus on these modes.

The solution of eq. (\ref{zereq}) 
can be constructed by the Green-function technique as follows.
We first integrate the interior  equations (\ref{poleq}) 
from $r=0$ to  $r=R$,
where we compute the Zerilli function 
\op Z_{l m}(R)\cl as given in eq. (\ref{zerfun}), and its
first derivative. 
We then construct, by numerical integration, 
a solution of the homogenous Zerilli
equation, $Z_{l m}^1$,  which satisfies the boundary condition
\op Z_{l m}^1(R)=Z_{l m}(R);\cl
at radial infinity $Z^1$ behaves as a superposition of ingoing and
outgoing waves,
\op
Z_{l m}^1\sim A_{l m}^{in}e^{-i\omega r_*}+A_{l m}^{out}e^{+i\omega r_*}.
\cl
Let us consider a second solution of the homogenous Zerilli equation, 
$Z_{l m}^2$, which behaves as a pure outgoing wave 
at infinity, i.e.
\op
Z_{l m}^2\sim e^{+i\omega r_\star}.
\cl
The general solution of the non-homogeneous Zerilli equation (\ref{zereq}),
which satisfies the physical requirement of pure outgoing radiation at
infinity,
\op
Z_{l m}(r\to \infty)\sim Z_{l m}^{out}(\omega)
e^{+i\omega r_\star},
\cl and the matching condition at the surface,
can be written in terms of these two functions 
\be
Z_{l m}=\f{1}{W_{l m}}\left\{Z_{l m}^2\int_{R}^{r_*} Z_{l m}^1S_{l m}  dr_\star
-Z_{l m}^1\int_{\infty}^{r_*}Z_{l m}^2S_{l m} dr_\star \right\}
\ee
where $W_{l m}$ is the wronskian of the two solutions
\be
W_{l m}\equiv \left(Z_{l m}^1 Z^2_{{l m},r_\star}-
Z_{l m}^2 Z^1_{{l m},r_\star}\right)= 2 i\omega A_{l m}^{in}.
\ee
The amplitude of the Zerilli function at infinity therefore is:
\be\label{Zout}
Z_{l m}^{out}(\omega)=\f{1}{2i\omega A_{l m}^{in}}\int_{R}^{\infty}{Z_{l
m}^1S_{l m} dr_\star}.
\ee
Because of the form of the source term (\ref{source}), (\ref{source1}), this
amplitude can be written as
\[
Z_{l m}^{out}\equiv M_p \bar{Z}_{l m}^{out}(m\omega_k,R_0)\delta(\omega-m\omega_k),
\]
It should be noted that
if we restrict our analysis to $l=2$, the source is non zero
only for $m=\pm 2$. 
Finally, the amplitude of the outgoing gravitational radiation at infinity can
be computed by using the relation between $Z^{out}$ and the two
wave components $h_\pm$  in the radiative gauge\cite{zerilli}
as follows
\beq
\label{amplitude}
\left[
\bar{h}_+(2\omega_k,r)+i 
\bar{h}_-(2\omega_k,r)\right]\\\nn
=-
\f{M_p}{r} \sum_{ m} 2\sqrt{n(n+1)} 
\bar{Z}_{2 m}^{out}(m\omega_k,R_0) ~ {_{2}}Y_{2 m}(\vartheta,\varphi),
\eeq
where ${_{2}}Y_{l m}$ is the $s=2$ spin-weighted spherical harmonic.
As in  section III, we shall define the relativistic 
characteristic amplitude, $h_R^{l=2}$, 
to be compared with $h_Q$, by
\be
\label{hR}
h_R^{l=2}
\equiv \sqrt{\f{2}{3}}\sqrt{\f{1}{4\pi}\int{d\Omega
\left[\left|\bar{h}_+\right|^2
+\left|\bar{h}_-\right|^2\right]}}=
\f{M_p}{r} \f{2\sqrt{n(n+1)}
\left|\bar{Z}_{22}^{out}(2\omega_k, R_0)\right|}{\sqrt{3\pi}},
\ee
where we have used the orthonormality of the
spin-weighted harmonics, and, in summing over $m$, the symmetry property
$\left|\bar{Z}_{22}^{out}\right|^2=\left|\bar{Z}_{2,-2}^{out}\right|^2$.

We have computed  $h_R^{l=2}\cl 
assuming that the planet moves on a circular orbit of radius
$R_0=R^i_{res}+\Delta R,$  where $R^i_{res}$ is the orbit corresponding to
the  excitation of the  mode $g_i$, for the modes allowed by the Roche
lobe analysis.
We find that, as the planet approaches the resonant orbit, 
$h_R^{l=2}\cl grows very sharply.
It is instructive to plot  the ratio $h_R^{l=2}(R_0)/h_Q(R^i_{res})$
as a function of \op \Delta R,\cl to see  how much the amplitude of the
emitted wave grows with respect to the quadrupole emission, because 
of the excitation of a g-mode.
In figure 1 the logarithm of this ratio is plotted for the modes g$_4$,
g$_7$, and g$_{10}$ as a function of \op\log\Delta R,$ 
showing a power-law behaviour nearly independent of the 
order of the mode.
It should be stressed that $h_R^{l=2}(R_0)/h_Q(R^i_{res})$ 
is independent of the mass
of the planet, but depends, of course, on the selected stellar model.
From figure 1 we deduce that, in principle, as the planet approaches
a resonant orbit  the amplitude of the emitted wave may 
become significantly higher than that emitted because of the orbital
motion ($h_Q$). Thus a relevant question to answer is how long 
can a planet move on an orbit close to a resonance, 
before radiation reaction effects move it off resonance.
This issue will be discussed in the following section.
\section{The effect of radiation reaction}

The loss of energy in gravitational waves causes a shrinking 
of the orbit of a planetary system, and the efficiency of this process
increases as the planet approaches a resonant orbit.
We shall now compute the time a planet takes to move from an orbit of radius
$R_0=R^i_{res}+\Delta R,$  where  the amplification  factor
$h_R/h_Q$ has some  assigned value,
to the resonant orbit $R^i_{res}$, because of
radiation reaction effects.
This timescale will indicate whether a planet can stay in the
resonant region long enough to be possibly observed.
On the assumption that the timescale over which the orbital radius evolves
is much longer than the orbital period (adiabatic approximation),
the  orbital shrinking  can be computed  from  the energy conservation law
\be\label{energycons}
M_p\left\langle\f{dE}{dt}\right\rangle
+\left\langle\f{dE_{GW}}{dt}\right\rangle=0,
\ee
where $E$ is the energy per unit mass of the planet 
as given in (\ref{energyandmom}), and 
$\left\langle\f{dE_{GW}}{dt}\right\rangle$ is
the energy emitted in gravitational waves, which can be computed
in terms of the wave amplitude  (\ref{Zout}) as follows
(cfr. e.g.\cite{tanaka})
\be
\label{engrav}
\left\langle \f{dE_{GW}}{dt} \right\rangle
=\lim_{T\to\infty}\f{E_{GW}}{T}=
\lim_{T\to\infty}\f{1}{T}\int \f{dE_{GW}}{d\omega} d\omega=
\sum_{l m}\f{(l-1)l(l+1)(l+2)}{32\pi}
(m\omega_k  M_p)^2
\left|\bar{Z}^{out}_{\ell m}(m\omega_k)\right|^2.
\ee
Since 
$\left\langle\f{dE}{dt}\right\rangle=
\left\langle\f{dR_0}{dt}\right\rangle/\left\langle\f{dR_0}{dE}\right\rangle$,
using  eq. (\ref{energyandmom}) and eq. (\ref{engrav}),
eq. (\ref{energycons}) gives
\be
\left\langle\f{dR_0}{dt}\right\rangle=-\f{2 R_0^2}{M_p M_\star}
\f{(1-3M_\star/R_0)^{3/2}}{(1-6M_\star/R_0)}
\left\langle\f{dE_{GW}}{dt}\right\rangle,
\ee
from which the time needed  for the planet to reach the resonant orbit
can be computed
\be
\label{deltat}
\Delta T=
-\f{M_p M_\star}{2}
\int_{R^i_{res}+\Delta R}^{R^i_{res}}
\f{(1-6M_\star/R_0)}{\left\langle\f{dE_{GW}}{dt}\right\rangle 
(1-3M_\star/R_0)^{3/2}}\f{dR_0}{R_0^2}.
\ee
It should be noted that since $\left\langle\f{dE_{GW}}{dt}\right\rangle$
is proportional to $M_p^2$, \op \Delta T\cl is longer for  smaller
planets.
We have computed $\Delta T$ for the three companions $M_E,$ $M_J$ and
$M_{BD},$  by the following steps:\\
- For each g-mode allowed by the Roche lobe analysis we find the radius
of the  resonant orbit  $R^i_{res}$.\\
- We assume that each companion
orbits the star at a distance $R_0=R^i_{res}+\Delta R$ such that 
the amplification
factor \op A= h_R(R_0)/h_Q(R^i_{res})\cl has an assigned value,
and compute the corresponding
energy radiated in gravitational waves,
$\left\langle\f{dE_{GW}}{dt}\right\rangle$.\\
- We compute $\Delta T$, which tells us how long can the
companion orbit the star in the resonant region between $R_0$ and 
$R^i_{res}$, emitting a wave of
amplitude higher than $A~ h_Q(R^i_{res})$.\\
The results are summarized in table \ref{times}.
These data have to be used together with  those  in table 
\ref{amplitudes} as follows.
Consider for instance a planet like the Earth, orbiting its star 
on an orbit resonant with the mode g$_4$. 
According to the quadrupole formalism, which does 
not consider changes in the internal structure of the star
and therefore does not include the
resonant contributions to the emitted radiation,
it would emit a signal of amplitude $h_Q(g_4)=3\cdot
10^{-26},\cl at a frequency $\nu_{GW}=1.1\cdot 10^{-4}$ Hz
(table \ref{amplitudes}, first and last row, respectively).
The data of table \ref{times}, which include the resonant contribution,
indicate  that before reaching the resonant orbit $R^4_{res}$,
the Earth-like planet would orbit in a region  of thickness
 $\Delta R =8.7$ km
slowly spiralling in, emitting waves with amplitude
$h_R > 10 h_Q(g_4)=3\cdot 10^{-25},\cl for a time interval 
of $8.7\cdot 10^6$ years, and that while spanning the smaller radial
region  $\Delta R =1.3$ km,
the emitted wave  would reach an amplitude 
$h_R > 50 h_Q(g_4)=1.5\cdot 10^{-24},\cl for a time interval 
of $4.7\cdot 10^4$ years.

A jovian planet, on the other hand, could only excite modes of order 
$n=10$ or higher, which would correspond to a resonant frequency of 
$\nu_{GW}=5.7\cdot 10^{-5}$ Hz and a
gravitational wave amplitude greater than $6\cdot 10^{-23}$ for  520
years ($\Delta R =42$ m), or greater than 
$3\cdot 10^{-22}$ for $\sim 3$ years ($\Delta R =6$m).
From these data we see that the higher  the order of the mode, the more
difficult it is to excite it, because the region where the resonant
effects become significant gets narrower and the planet transits through
it for a shorter time.

Much more interesting are the data for a brown dwarf companion.
In this case, for instance, the region which would correspond 
to the resonant excitation of the mode g$_4$ ($\Delta R=8.7$ km),
with a wave amplitude greater than $3.9\cdot 10^{-21}$, would  be
spanned in $690$ years,
whereas the emitted wave would have an amplitude greater than
$\sim 2\cdot 10^{-20}$ ($\Delta R=1.3$ km)
over a time interval of $\sim 4$ years.

\section{Concluding Remarks}
The study carried out in this paper has been motivated by the recent 
discovery of several extrasolar planetary systems very close to Earth,
with companions orbiting the host star much closer than previously expected.
This discovery suggests the possibility that a companion may be found orbiting
at such close distance  to excite one of the g-modes of the star, and 
a Roche-lobe analysis indicates that this may be possible, especially 
if the companion is a brown dwarf.
The frequency of a g-mode in a sun-like star is of the order of a 
few digits in $10^{-4}$ Hz, and would be in the bandwith of the spaceborn
interferometer LISA. 
Thus, our aim was to understand how much such resonant process may enhance 
the emission of gravitational waves of the system with respect to the
radiation it emits because of the time varying quadrupole moment
associated to the orbital motion.
This enhancement has been  evaluated by integrating the equations that
describe the perturbation the companion induces on the star,
and  computing the amplitude of the emitted wave when the planet
moves close to an orbit corresponding to
the resonant excitation of a g-mode.
The amplitude increases sharply as the orbit approaches the resonant
one; however, in order to get a chance to observe one of these signals
emitted by a system in our vicinity, it is also needed 
that  it stays above a certain threshold long enough.
For this reason we have also computed the effects of
radiation reaction on the shrinking of the orbit and the associated
timescales. 
The main information we extract from  our  study  is that only close brown
dwarf companions may produce signals of some relevance;  a brown dwarf
of  40 Jovian masses, orbiting a star located at a distance 
of 10 pc from Earth, if close to an orbit resonant with the mode $g_4$,
may emit radiation of amplitude 
greater than $\sim 2\cdot 10^{-20}$ at a frequency
of the order of $\sim  10^{-4}$ Hz,  spanning a region 
$\Delta R=1.3$ km over a time interval of $\sim 4$ years.
Of course the amplitude of the wave increases with the mass of the brown
dwarf  and decreases with the distance,  and
it will be interesting to see whether astronomical observations 
will identify any  such system in our vicinity in the future.

As a last remark, we would like to mention that
things would not be significantly different had we considered more
realistic models of solar-type stars, such as those discussed in
\cite{christensen},\cite{dirty}; they predict oscillation frequencies
that are  a little higher than those of our simple polytropic
model; however, our frequencies
are  at least qualitatively correct to give order-of-magnitude 
estimates for the effects of resonant excitation of the stellar modes
and  for the excitation timescales.


\newpage
\begin{table}
\centering
\caption{
The minimum mean density,  $\rho_{min}/\rho_\odot$,
that a planet should have in order to be allowed to move on an orbit which
corresponds to the excitation of a g-mode,
without being disrupted by tidal interactions.
The values of $\rho_{min}/\rho_\odot$ are tabulated  for
three planets with mass
equal to that of the Earth ($M_E$), of Jupiter ($M_J$) and
of a brown dwarf with $M_{BD}=40~M_J$.
}
\vskip 10pt
\begin{tabular}{@{}lllllllllllll@{}}
                         &\multicolumn{11}{c}{$\rho_{min}/\rho_\odot$} \\
\hline
&
& $g_{1}$ & $g_{2}$ & $g_{3}$ & $g_{4}$ & $g_{5}$
& $g_{6}$ & $g_{7}$ & $g_{8}$ & $g_{9}$ & $g_{10}$ \\
\hline
&$M_{E}$ 
&12.5 &7.21 &4.64 & 3.24 & 2.38
& 1.83 & 1.45 & 1.17 & 0.97 & 0.82 \\
\hline
&$M_{J}$
&13.0 &7.49 &4.82 &3.37 &2.48
&1.90 &1.51 &1.22 &1.01 & 0.85 \\
\hline
&$M_{BD}$
&-    &-    &-    & 3.68 & 2.71
& 2.08 & 1.65 & 1.34 & 1.11 & 0.93 \\
\end{tabular}
\label{lobi}
\end{table} 
\begin{table}
\centering
\caption{ The amplitude of the gravitational signal emitted
when  the three companions considered in table 1
move on a circular orbit
of radius $R^i_{res}$, such that the condition of resonant excitation of a
g-mode  is satisfied,
is computed by the quadrupole formalism for the modes allowed  by the
Roche-lobe analysis.
The planetary systems are assumed to be at a distance of
$10$ pc from Earth.
In the last line  the emission frequencies $\nu_{GW}$ are given.
}
\begin{tabular}{@{}lllllllll@{}}
                         &\multicolumn{8}{c}{$h_Q(R^i_{res})$} \\
\hline
&
&  $g_{4}$ & $g_{5}$
& $g_{6}$ & $g_{7}$ & $g_{8}$ & $g_{9}$ & $g_{10}$ \\
\hline
&$M_E$
& $3.0\cdot 10^{-26}$ & $2.7\cdot 10^{-26}$ & $2.5\cdot 10^{-26}$
& $2.3\cdot 10^{-26}$ & $2.2\cdot 10^{-26}$
& $2.0\cdot 10^{-26}$ & $1.9\cdot 10^{-26}$  \\
\hline
&$M_{BD}$
& $3.9\cdot 10^{-22}$ & $3.5\cdot 10^{-22}$ & $3.2\cdot 10^{-22}$
& $3.0\cdot 10^{-22}$ & $2.8\cdot 10^{-22}$
& $2.6\cdot 10^{-22}$ & $2.4\cdot 10^{-22}$  \\
\hline
&$M_J$
& - &  - &  - &  - &  - &  - & $6.1\cdot 10^{-24}$\\
\hline
&$\nu_{GW}$($\mu$Hz)
& $112.5$ & $96.6$ & $84.7$ & $75.4$ & $67.9$
& $61.8$ & $56.7$  \\
\end{tabular}
\label{amplitudes}
\end{table}

\begin{table}
\centering
\caption{
In column 2 we give the frequency, \op \nu_{GW},\cl  of the wave
emitted when a
companion moves around the host star on an orbit resonant with
a g-mode allowed by the Roche lobe analysis (see text).
Because of resonant effects, the amplitude of the wave is amplified by
a factor greater than $A$ (column 3) when the companion spans a radial 
region of thickness \op \Delta R\cl (column 4), which
is the same for all planets.
In the last three columns we give  the time interval  $\Delta T$ needed for
the three companions to  span the region $\Delta R$ and 
reach the resonance, because of radiation reaction effects.
}
\begin{tabular}{@{}cllllll@{}}
\\
\multicolumn{1}{c} 
{Mode}&$\nu_{GW}$ ($\mu$Hz)& $A$&$\Delta
R$(m)&$\Delta T_E$(yrs)&$T_{BD}$(yrs)&$T_{J}$(yrs)\\
\hline
g$_4$ &$112.5$  &$10$   &8740   &$8.7\cdot10^6$ &$6.9\cdot10^2$         &-\\
      &  &$50$   &1326   &$4.7\cdot10^4$ &$3.7$                  &-\\
\hline
g$_5$ &$96.6$  &$10$   &3166   &$4.3\cdot10^6$ &$3.4\cdot10^2$         &-\\
      &  &$50$   &481    &$2.3\cdot10^4$ &$1.8$                  &-\\
\hline
g$_6$ &$84.7$   &$10$   &1234   &$2.2\cdot10^6$ &$1.7\cdot10^2$         &-\\

      &  &$50$   &187    &$1.2\cdot10^4$ &$9.2\cdot10^{-1}$      &-\\
\hline
g$_7$  &$75.4$ &$10$   &506    &$1.1\cdot10^6$ &$89$                   &-\\
      &  &$50$   &77     &$6.0\cdot10^3$ &$4.7\cdot10^{-1}$      &-\\
\hline
g$_8$ &$67.9$  &$10$   &216    &$5.9\cdot10^5$ &$47$                   &-\\
      &  &$50$   &33     &$3.1\cdot10^3$ &$2.4\cdot10^{-1}$      &-\\
\hline
g$_9$ &$61.8$  &$10$   &94     &$3.1\cdot10^5$ &$24$                   &-\\
      &  &$50$   &14     &$1.7\cdot10^3$ &$1.3\cdot10^{-1}$      &-\\
\hline
g$_{10}$ &$56.7$ &$10$   &42     &$1.7\cdot10^5$ &$13$      &$5.2\cdot
10^2$
\\
      &  &$50$   &6      &$9.2\cdot10^2$ &$7.2\cdot10^{-2}$      &$2.9$\\
\end{tabular}
\label{times}
\end{table}

\newpage
\begin{figure}
\centerline{\hbox{\psfig{figure=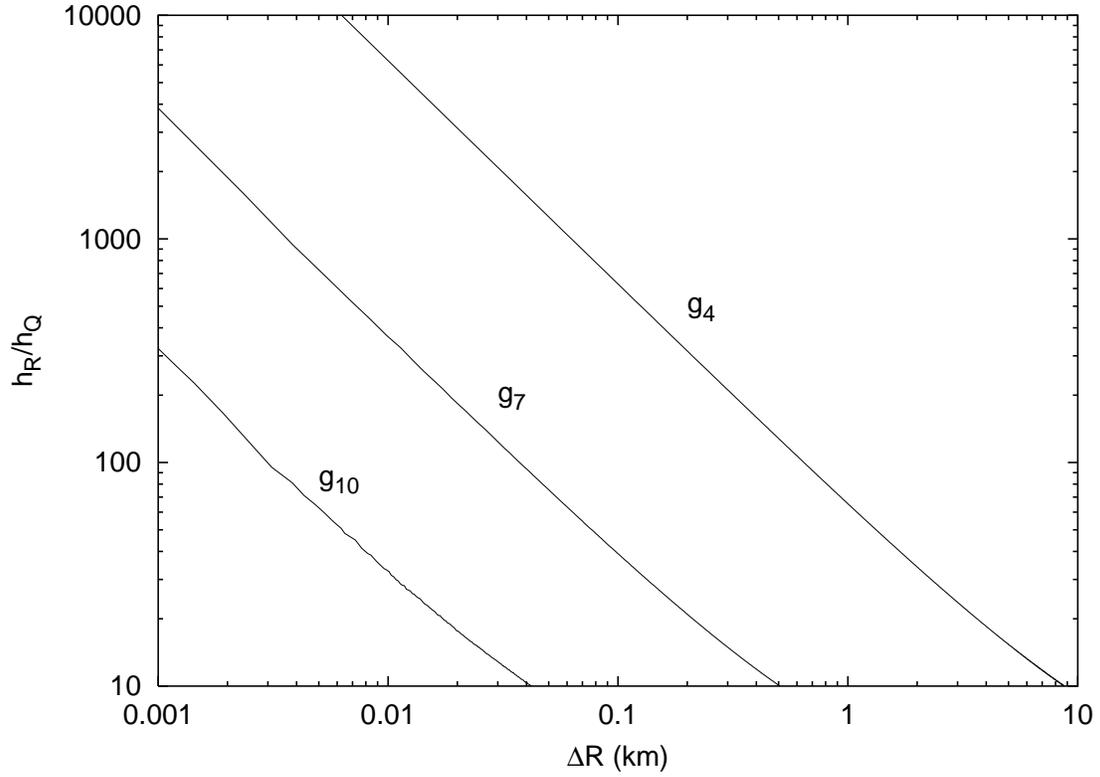,angle=270,width=15cm} }}
\vskip 4pt
\caption{
The logarithm of the ratio $h_R^{l=2}(R^i_{res}+\Delta R)/h_Q(R^i_{res})$
is plotted as a function of $ \Delta R$ for the modes g$_4$,
g$_7$, and g$_{10}$ (see text).}
\label{fig1}
\end{figure}

\end{document}